\documentclass{ws-rv9x6}
\usepackage{ws-rv-van}     
\usepackage{caption}
\usepackage{subcaption}

\usepackage[pdftex]{hyperref}
\hypersetup{colorlinks=true,
            urlcolor=blue,
            citecolor=blue,
            pdfstartview={FitH},
            pdftitle={Search in random media with Lévy flights},
            pdfauthor={Erol Gelenbe, Omer H. Abdelrahman}}

\makeindex
\newtheorem{result}[theorem]{Result}
\DeclareMathOperator{\erfc}{erfc}

\begin{document}

\chapter[Search in Random Media]{Search in Random Media with L\'evy Flights}\label{ra_ch1}

\author[E. Gelenbe and O. H. Abdelrahman]{Erol Gelenbe and Omer H. Abdelrahman}

\address{Department of Electrical \& Electronic Engineering,\\
Imperial College London, SW7 2BT, UK \\
\{e.gelenbe,o.abd06\}@imperial.ac.uk}
\body

\section{Introduction and Motivation}

Our work on the subject of search has been motivated by a variety of applications in engineering and technology, and it differs from the physicists'
motivation for such problems, e.g. \cite{Benichou2004,Benichou2007,Evans2011b}, or the motivation provided by biochemistry \cite{Berg1981,Eliazar2008}, and in theoretical biology
\cite{Tilch99a,Oshanin2009,Rojo2010,Redner1} .

One of our reasons for a more fundamental approach to search comes from
the area of traffic routing in wired networks \cite{Gelenbe_ICTAI1999,CPN-Patent,Gelenbe2009} where each end user may send out many search or
Smart Packets to seek out paths for the traffic it needs to forward, so that the paths that are used will satisfy useful quality of service (QoS) properties such as minimising delay or packet loss.
A similar question arises, with even more acuity, when the networks are ad-hoc and wireless, and the need to reduce uncertainty regarding possible paths is
even greater \cite{Gelenbe2006,Gelenbe2007,Shakkottai05}. A distinct but related motivation for our work
has been the search for network paths that avoid ongoing Cyberattacks \cite{DoS}.

Another important motivation for our work comes from the field of explosive mine detection and removal on which one of us worked in the past \cite{SearchMines,Mines},
and in such circumstances it is sensible to see how the natural world deals with such challenges \cite{Gelenbe1997,Natural}. The related fields of tactical operations \cite{Virtual}, and emergency operations and management \cite{Filippoupolitis2008_AmbiSys,Dimakis_2010_EvacuationSimulator,ICCCN2012,CAMWA2012,Desmet2} also raise many interesting problems
related to the search for targets \cite{SearchMines}, objects such as landmines \cite{Mines}, and victims, as part of an organised response to an incident or an emergency.

On the other hand, one of us first introduced L\'{e}vy flights in diffusion models in the 1970's,  motivated by the performance analysis of computer systems \cite{Gelenbe1975} to deal with continuous approximations to discrete queueing models under light traffic conditions, whereas the usual approach to
such approximations for queues have focused on heavy-traffic conditions. Here, the instantaneous jump or L\'{e}vy flight was used to represent a jump from the origin (the empty state) of the diffusion process, to the value $+1$ representing the arrival of a customer to an empty queue. When the diffusion model was used to model the waiting time process of the queue, the instantaneous jump represented the service time of the first customer arriving to an empty queue \cite{Gelenbe1979} including non-Markovian multiple-phase models. This analysis then gave rise to a call admission control (CAC) algorithm for asynchronous transfer model (ATM) networks \cite{Gelenbe1996} which resulted in a patented CAC scheme for ATM networks \cite{GelenbeCAC-Patent}.

In the sequel we review some of our work regarding search, focusing on the L\'evy flight models, including the issues of time, energy, inhomogeneous media and security.
In Section \ref{avg-comput} we develop the basic mathematical model representing $N$ searchers that proceed independently, and which can be affected
by destruction or loss, and time-outs. A time-out is a length of time such that if a searcher has not been successful after the time-out elapses, the organiser of the search simply aborts that particular searcher and some time later replaces it with a new one. In related approaches \cite{Mejia2011}, the authors study the fluctuations, from one trajectory to another, of the first passage time of $N$ independent searchers.

In Section \ref{coupled} we present the system of $N$ coupled diffusion equations,
together with the discrete probability equations representing losses and time-outs, and subsequent restarts, and their analytical solution in steady-state
is obtained in Section \ref{stat}.

An approach based on Laplace-Stieltjes transform of the search time is then developed in Section \ref{trans-comput} for the case where the success of the search requires that $k$ out of the $N$ searchers be successful. Section \ref{asymptote1} then considers the problem of estimating the number $N$ of serachers that are needed so that $k$ out of these $N$ searchers will be successful in time $B$, provided $B$ is large. Then, computing the amount of energy needed to conduct a successful search is discussed
in Section \ref{energy}.

All the previous analysis assumed that the search space has homogeneous and time-invariant characteristics. Our work on non-homogeneous media is actually motivated by
security models where, for instance, as the searchers get closer to the object they seek, the search can become more difficult and searchers may be more frequently destroyed. This is certainly the case where countermeasures are in place to protect or defend the immediate surroundings of (say) a network node that is being defended against Cyberattacks. Thus in Section \ref{non-hom} we present an iterative numerical solution approach that allows us to analyse the search time and the energy needed to find an object when the search space is non-homogeneous. We then apply this approach in Section \ref{protect} to the analysis of a {\em region of protection} around an object that is sought out by a searcher. Finally in Section \ref{phase} we show that there is a form of phase transition that occurs when the time-out is varied in relation to the
rate at which searchers are lost or destroyed; thus it is possible to reach a situation where the searcher (or attacker) will always be successful against the defender that destroys the attacker in the proximity of the object being sought out (and defended). Some suggestions for further research are made in the final section of this paper.

\section{Modelling the Search Process}\label{avg-comput}

Consider $N$ independent searchers sent out simultaneously in the quest for the same object. Let $Z_i(t)$ denote the $i$-th searcher's {\em distance} from its destination at time $t\geq 0$ with $Z_i(0)=D$, and denote its search time by $T_i = \inf\{t: Z_i(t)=0 \}$. Let $T_{1,N}\leq T_{2,N}\leq \cdots \leq T_{N,N}$ be the variables $T_i$ rearranged in ascending order, i.e. the corresponding order statistics; this section deals with the computation of $E[T_{1,N}]$ \cite{Gelenbe2010}. The state of the searcher at time $t\geq 0$ is $s_i(t)$ which can take one of the values $\{{\bf S_i,L_i,W_i,P}\}$ defined as follows:\medskip

\noindent $\bullet$ $\bf S_i$: If the $i$-th searcher is searching and its distance from the destination is $Z_i(t)>0$. We denote the probability density function (pdf) of the distance $Z_i(t)$ by $f_i(z_i,t)dz_i=P[z_i<Z_i(t)\leq z_i+dz_i, ~s_i(t)={\bf S_i}]$.\medskip

\noindent $\bullet$ $\bf W_i$: The $i$-th searcher's life-span has ended, and so has its search. Note that this may have happened because it was destroyed or became lost, but this becomes known to the source via the time-out which is exponentially distributed with parameter $r$. After an additional exponentially distributed delay of parameter $\mu$, it is replaced at the source by a new searcher with the same identity. We write $W_i(t)=P[s_i(t)={\bf W_i}]$.\medskip

\noindent $\bullet$ $\bf L_i$: The $i$-th searcher has been destroyed or lost, and its search is ended; for small $\Delta t$ and $Z_i(t)=z>0$, this happens with a probability $\lambda(z) \Delta t + o(\Delta t)$, where $\lambda(z)\geq 0$ is the destruction rate at distance $z$. The time spent in this state is exponentially distributed with the same parameter $r$ as the life-span since the source realises that the searcher is lost or destroyed via the life-span effect. At the end of this exponentially distributed time, the searcher is handled just as if it has ``died''. We write $L_i(t)=P[s_i(t)={\bf L_i}]$.\medskip

\noindent $\bullet$ $\bf P$: One of the searchers has found the object being sought; the search process stops for all searchers, including the ones who are lost or dead.  Notice that $\bf P$ is a synchronised state for all of the searchers. After one time unit, the search process starts again as before at the source with $N$ searchers being sent out. We write $P(t)=P[s_i(t)={\bf P}]$.\medskip

\begin{figure}[t]\centering%
   \includegraphics[width=0.98\textwidth]{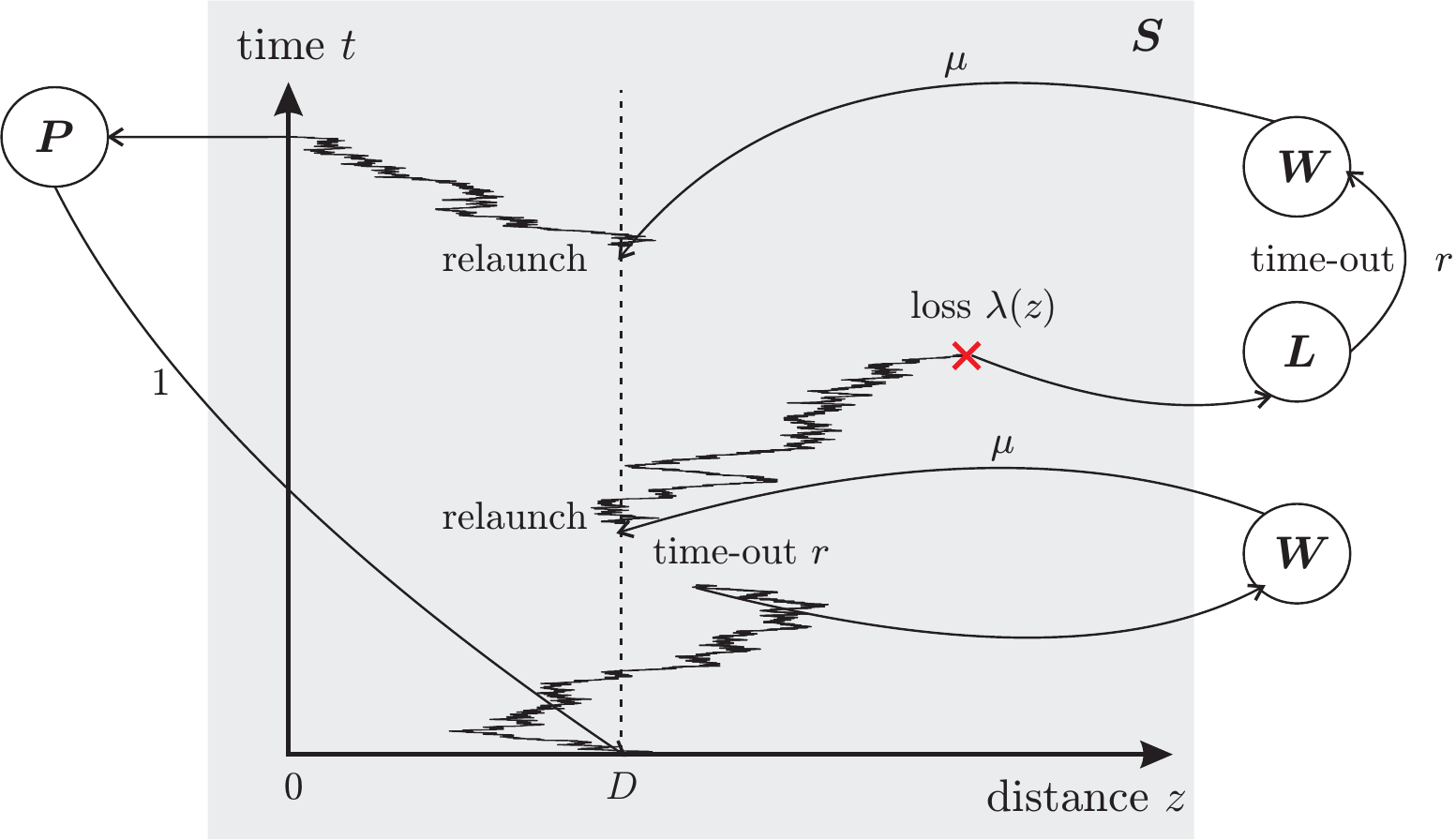}
   \caption{An instance of the search process with $N=1$ where the search is relaunched twice, due to timeout and destruction, before the object is finally found.} \label{diagram}
\end{figure}

Notice that the above process repeats itself indefinitely, and $E[T_{1,N}]$ is the average time that it takes from any successive start of the search until the first instance when state $\bf P$ is reached again. Strictly speaking, the L\'evy flight takes place only in this last instance, since diffusion process is reset, or jumps  to the distance $D$ for all of the searchers. However in a more abstract sense, whenever the search process of any one of the searchers leaves the diffusion in order to join another discrete probability variable, i.e. such as $L_i(t)$ or $W_i(t)$, we have a form of L\'evy flight as an instantaneous jump away from the diffusion and into a discrete set of states, before the process restarts at a later time. A schematic representation of the search process for $N=1$ is presented in Fig.~\ref{diagram}.

Let $P(t)$ be the probability that the model we have just described is in state $\bf P$ at time $t\geq 0$, and let $P=\lim_{t\to\infty}P(t)$. Then:
\begin{equation}
P=\frac{1}{1 + E[T_{1,N}]},\quad E[T_{1,N}] =\frac{1-P}{P}. \label{T1N}
\end{equation}
We assume that a searcher consumes energy only while it is {\em moving} (i.e., in state $\bf S_i$) at one energy unit per unit time, which is equivalent to stating that the energy consumed by a searcher is the same as the time it spends in motion, to the exclusion of epochs spent waiting for a time-out to relaunch it. A real number $b(z)$ represents the average rate of change over time $[t,t+\Delta t)$ of the searcher's distance to the destination, and the variance of the distance travelled by the searcher in that time interval is $c(z)\Delta t$, $c(z)\geq 0$:
\begin{align}
b(z)=&\lim_{\Delta t\rightarrow 0}\frac{E[Z(t+\Delta t)-Z(t) |Z(t)=z]}{\Delta t}, \nonumber \\
c(z)=&\lim_{\Delta t\rightarrow 0}\frac{E[(Z(t+\Delta t)-Z(t))^2]-(E[Z(t+\Delta t)-Z(t)])^2| Z(t)=z]}{\Delta t}. \nonumber
\end{align}
While $b(z)<0$ is the favourable case where the searcher on average gets closer to the destination with time, we may also have cases of interest with $b(z)>0$, which means that the searcher on average moves {\em away} from the object of interest, for instance because intermediate locations provide wrong information on average, or it lacks information altogether when $b(z)=0$. It was shown that \cite{Gelenbe2006,Gelenbe2007} in a homogeneous medium, even if $b\geq 0$ it is possible to have a travel time to destination which is finite on average.

\subsection{The System of Coupled Equations} \label{coupled}

We now express the process $\{s_i(t): t\geq 0\}$ in terms of a system of equations describing a mixed continuous space (diffusion) supplemented by a discrete space random process \cite{Einstein1926,Gelenbe1975,Gelenbe1979,Gelenbe1996},
where the discrete part in this case describes the reset process. Note that in \cite{Gelenbe1975,Gelenbe1979,Gelenbe1996} the discrete portion is related to transitions to internal states
where the transition time is non-exponential so that the process is non-Markovian. We first write the equations that the probability density function $f_i(z_i,t)$, $z_i>0$,  and the probability masses $L_i(t)$, $W_i(t)$ and $P(t)$, $t\geq 0$ will satisfy. We represent the interaction between the diffusion processes using a parameter $a_i(t)$, $1\leq i\leq N$ in the following manner: $a_i(t)$ is the total rate of attraction exerted at time $t$ by all other diffusion processes, on the $i$-th diffusion due to the fact that one of the other diffusions may have reached its absorbing barrier. The system of coupled differential and partial differential equations representing the search is:
\begin{align}\nonumber
\frac{\partial f_i(z_i,t)}{\partial t}=&\frac{1}{2}\frac{\partial^2[c(z_i)f_i(z_i,t)]}{\partial{z_i}^2}-\frac{\partial[b(z_i)f_i(z_i,t)]}{\partial z_i}\\ \label{one}
&  -(\lambda(z_i)+r+a_i(t))f_i(z_i,t)+[P(t)+\mu W_i(t)]\delta(z_i-D),
\end{align}
while
\begin{align}\label{two}
\frac{dL_i(t)}{dt}=&-(r+a_i(t))L_i(t)+  \int_{0^+}^\infty \lambda(z_i) f_i(z_i,t)dz_i,\\ \label{three}
\frac{dW_i(t)}{dt}=&-(\mu+a_i(t))W_i(t)+r(L_i(t)+\int_{0^+}^\infty f_i(z_i,t)dz_i),\\ \label{four}
\frac{dP(t)}{dt}  =& -P(t)+\sum_{i=1}^N\lim_{z_i\to 0^+}[\frac{1}{2}\frac{c(z_i)\partial f_i(z_i,t)}{\partial z_i}-b(z_i) f_i(z_i,t) ],
\end{align}
and
\begin{equation} \label{five}
a_i(t)=\sum_{j=1,j\neq i}^N\lim_{z_j\to 0^+}[\frac{c}{2}\frac{\partial f_j(z_i,t)}{\partial z_j}-bf_j(z_j,t)],
\end{equation}
which is the rate at which the $i$-th searcher is attracted to the origin, i.e. to finish its search, because any one of the {\em other} $N-1$ searchers has found the object being sought. This is reflected both in \eqref{one} and in the equations \eqref{two} and \eqref{three} where the searcher can be forced into the rest state from the ``lost'' state and the ``time-out before retransmission'' state, as well. We also see that we enter the loss state from any position $z_i>0$, and that a time-out can occur for a searcher that is in the lost state. Since the behaviour of all searchers when they are not in the rest state are independent, it follows that
the event that triggers the jump of searcher $i$ into the rest state does not depend on the prior state of searcher $i$ but on the state of the {\em other} searchers.
We also have that the sum of the probabilities is one:
\begin{equation}
1 = L_i(t) + W_i(t) + P(t) + {\int_{0^+}^\infty}f_i(z_i,t)dz_i. \label{total}
\end{equation}
Note again that these equations represent the system where, whenever any one searcher has reached the destination, all other searchers' progress is artificially stopped and re-started from the rest state. The purpose here is to compute $E[T_{1,N}]$ by constructing a synthetic ergodic process. In this section we assume that the medium in which the searchers move is homogeneous, i.e. $b(z)=b$, $c(z)=c$ and $\lambda(z)=\lambda$, while in Section \ref{non-hom} we deal with spatially non-homogeneous environments.

\subsection{The Stationary Solution} \label{stat}

Dropping the dependence on $i$ because all searchers are statistically identical, and writing the system of differential equations in steady-state yield:
\begin{align}\nonumber
0 ~=~& \frac{c}{2}\frac{d^2f(z)}{dz^2}-b\frac{df(z)}{dz}-(\lambda+r+a)f(z) +[P+\mu W]\delta(z-D),\\\nonumber
1 ~=~& P + \frac{\mu+r+a}{r} W,\\\nonumber
P ~=~& N\lim_{z\to 0^+}[\frac{c}{2}\frac{d f(z)}{dz}-b f(z) ],\\\label{steady-state}
a ~=~&(N-1)\lim_{z\to 0^+}[\frac{c}{2}\frac{\partial f(z)}{\partial z}-bf(z)].
\end{align}
The solution for $f(z)$ can be shown to take the following form:
\begin{equation}\label{fz}
f(z)=\left\{
\begin{array}{ll}
A [e^{u z }-  e^{v z }] ,& z\leq D, \\%
A [e^{(u-v)D}-1] e^{v z},& z\geq D,
\end{array}\right.
\end{equation}
where $A$ depends on $a$, and $u, v$ are, respectively, the positive and negative real roots of the characteristic polynomial of the differential equation \eqref{one}:
\begin{equation*}
u,v=\frac{b\pm \sqrt{b^2+2c(\lambda+r+a)}}{c}.
\end{equation*}
Solving the system of equations in \eqref{steady-state} with \eqref{fz} and \eqref{T1N} we obtain:
\begin{result}
The average time for one of $N$ searchers to find the object being sought is:
\begin{equation}\label{ET1N}
E[T_{1,N}] = \frac{\mu+r+a}{(\mu+a)(r+a)} \frac{e^{uD}-1}{N},
\end{equation}
where $a$ is given by the solution of the non-linear equation:
\begin{equation}
a = \frac{N-1}{N+\frac{\mu+r+a}{(\mu+a)(r+a)}[e^{uD}-1]}.
\end{equation}
The average energy consumed {\em until one of the searchers is successful} is:
\begin{equation}
E[J_{1,N}^-] = N (1+E[T_{1,N}])\int_{0^+}^\infty f(z)dz = \frac{e^{uD}-1}{\lambda+r+a}.
\end{equation}
\end{result}

\section{A Transform Approach when $k$ out of $N$ Searchers must be Successful}\label{trans-comput}

In this section we focus on the case where at least $k$ out of $N$ searchers must be successful \cite{Search-PhysRev2013}, and obtain approximate and asymptotic estimates for the search times $T_{k,N}$. Indeed, in hunting or foraging, search may take place as a group effort \cite{Gelenbe1997}, and $k$ out of $N$ members of a team must be successful in reaching the object. Another important motivation comes from communication networks \cite{Wang05} where packets can be encoded so that the information transmitted is correctly received when at least $k$ out of $N$ transmitted packets arrive at the destination, in the presence of possible packet loss. The energy consumed is computed (i) when the search is stopped as soon as the first $k$ searchers find the object, and (ii) when the remaining searchers (other than the first $k$ successful ones) continue their search till successful completion or until they are destroyed or stopped by their time-out. The latter case is relevant when the source cannot communicate with searchers or when successful searchers cannot communicate with their peers. We use a Laplace transform (LT) approach to compute the distribution of the search time and of the total energy consumption in the two aforementioned cases. Thus the state of a searcher at time $t$ is again $s(t)\in\{\bf S,W,L,P\}$ but unlike the recurrent approach of Section~\ref{avg-comput}, $\bf P$ is now an absorbing state.

\begin{figure}
        \centering
        \begin{subfigure}[t]{0.49\textwidth}\centering
                \includegraphics[width=\textwidth]{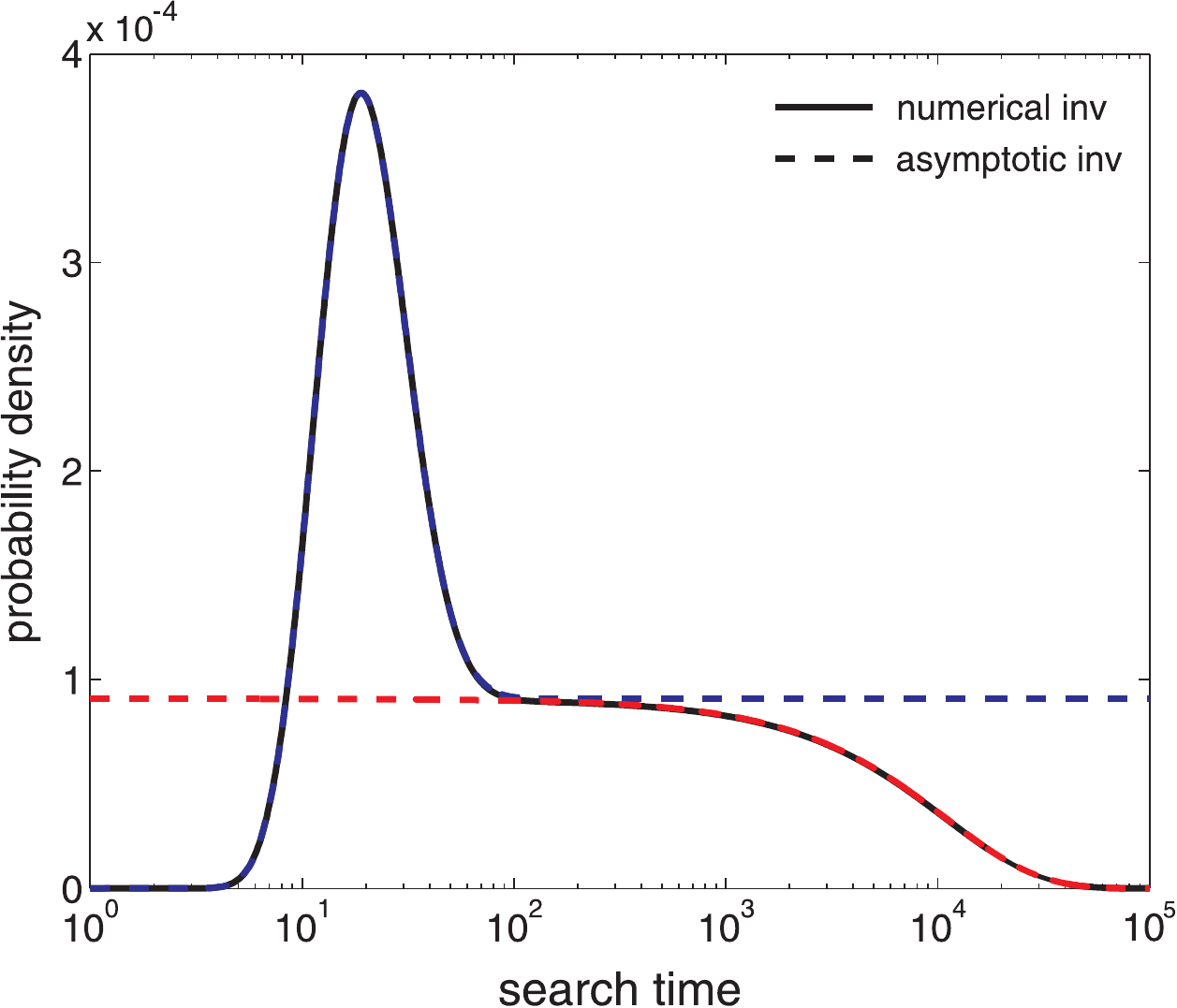}\caption{}\label{gh-a}
        \end{subfigure}~~
        \begin{subfigure}[t]{0.49\textwidth}\centering
                \includegraphics[width=\textwidth]{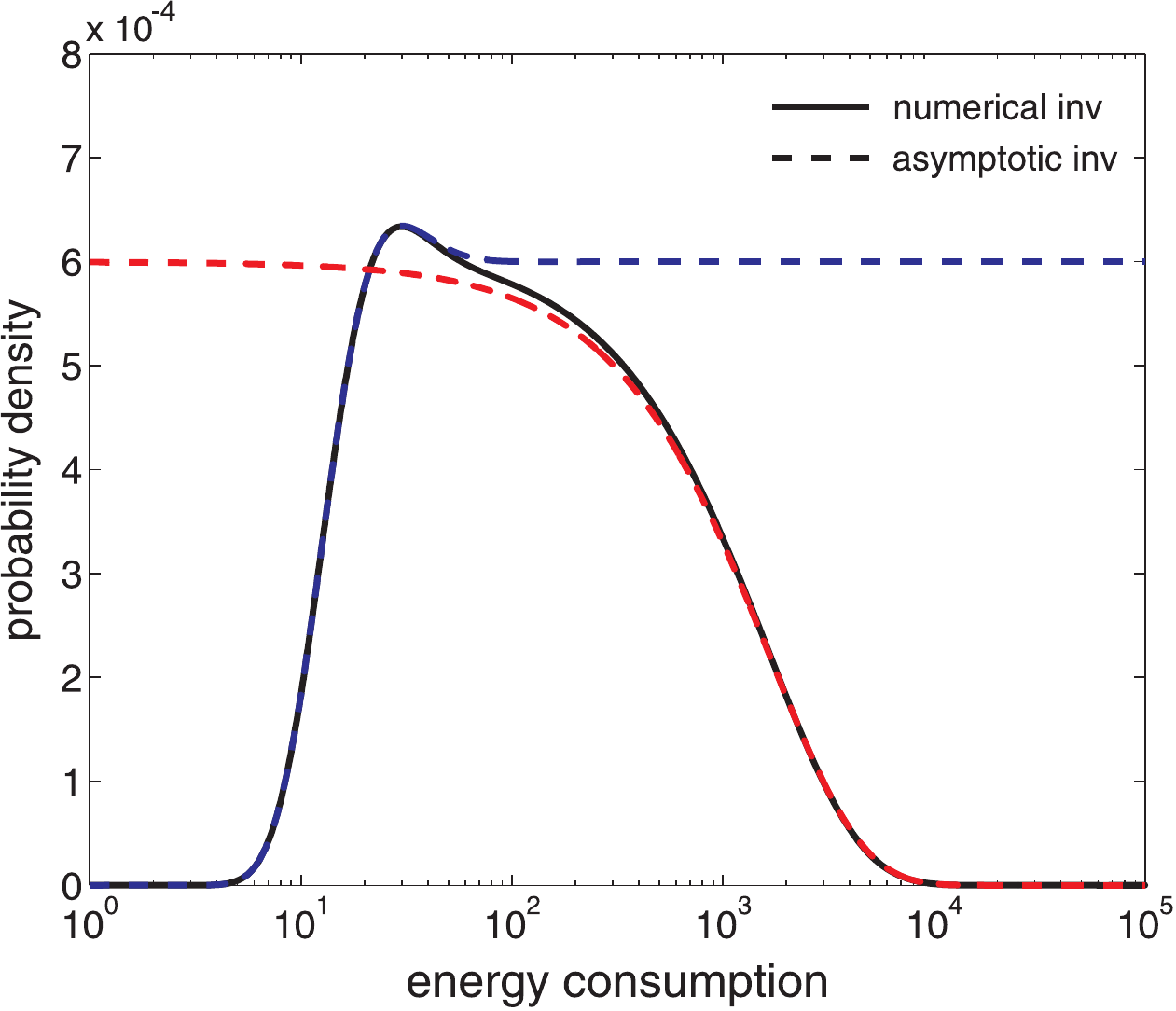}\caption{}\label{gh-b}
        \end{subfigure}
        \caption{The pdf of (a) search time $g(t)$  and (b) energy consumption $h(t)$ when $\lambda=0.05$, $b=0.1$,  $c=1$, $r=0.01$, $\mu=0.1$, and $D=10$. For this example that has a small positive value of $b$ and hence relatively high uncertainty in search direction, the pdf of the search time has a long tail which is apparent from the logarithmic scale on the horizontal axis.}\label{gh}
\end{figure}

The probability that a single searcher has reached the object by time $t$ is denoted by $G(t)\equiv \Pr[T\leq t]$ and its pdf $g(t)$. The time until return to the origin of a pure diffusion process starting at distance $D$ is \cite{Redner2001}:
\begin{align}\nonumber
g_0(t) =~& \frac{D}{\sqrt{2\pi c t^3}}~e^{-\frac{(D+bt)^2}{2ct}},\\\label{T0-sol}
G_0(t) =~& \frac{1}{2} \left[\erfc{\left(\frac{D+bt}{\sqrt{2ct}}\right)} +e^{-2bD/c}\erfc{\left(\frac{D-bt}{\sqrt{2ct}}\right)}\right],
\end{align}
where $\erfc(x)=\frac{2}{\sqrt{\pi}}\int_x^\infty e^{-y^2} dy$ and the subscript $0$ indicates that the quantity refers to a pure diffusion. The pdf of the searcher's distance $z$ to the destination at time $t$ is:
\begin{equation}\label{f0-sol}
f_0(z,t) = \frac{e^{-\frac{b^2t}{2c}} e^{-\frac{b}{c}(D-z)}}{\sqrt{2\pi  ct}} \big[e^{-\frac{(z-D)^2}{2ct}}-e^{-\frac{(z+D)^2}{2ct}}\big].
\end{equation}
Writing the LT of any $y(t)$ as  $\bar{y}(s)=\int_0^\infty y(t)e^{-st}dt$, the above quantities yield:
\begin{align}\nonumber
\bar{g}_0(s)   =& e^{-(b+\sqrt{b^2+2cs})D/c},\\\label{T0-LT}
\bar{f}_0(z,s) =& \frac{e^{b(z-D)/c}}{\sqrt{b^2+2cs}}\big[e^{-\frac{\sqrt{b^2+2cs}}{c}|z-D|}-e^{-\frac{\sqrt{b^2+2cs}}{c}(z+D)}\big].
\end{align}
In the model with loss and time-out, let $X$ and $Y$ be the mutually independent random variables representing the time to the next loss and the time to the next time-out, respectively, which are exponentially distributed with parameters $\lambda$ and $r$. Then $\gamma_\iota(t)$, the pdf of the duration of a search time until its first interruption is $\gamma_\iota(t)dt = \Pr[t\leq \min(X,Y) \leq t+dt ,~T_0>t]$ since $T_0$ is the total search time if there is no interruption and its pdf is given in \eqref{T0-sol}. Therefore:
\begin{equation}\label{gamma-interr}
\gamma_\iota(t) = (\lambda+r)e^{-(\lambda+r)t}[1-G_0(t)], \bar{\gamma}_\iota(s) = \frac{\lambda+r}{s+\lambda+r}~[1-\bar{g}_0(s+\lambda+r)].
\end{equation}
Search is interrupted randomly several times in this manner, and after each interruption it starts again at the origin after a further delay whose pdf is:
\begin{align*}
\psi(t)=&\frac{r}{\lambda+r}~ \mu e^{-\mu t} + \frac{\lambda}{\lambda+r} \int_0^t r e^{-ry}\mu e^{-\mu (t-y)}dy,\\
\bar{\psi}(s)=&\frac{s+\lambda+r}{\lambda+r}~\frac{\mu r}{(s+\mu)(s+r)}.
\end{align*}
The last and hence successful attempt at reaching the destination has a duration whose pdf $\gamma_d(t)dt=\Pr[t\leq T_0 \leq t+dt,~\min(X,Y)> t]$ or:
\begin{equation}\label{gamma-dest}
\gamma_d(t)=g_0(t)e^{-(\lambda+r)t},\quad \bar{\gamma}_d(s)=\bar{g}_0(s+\lambda+r).
\end{equation}
If the searcher is successful in locating the object being sought in its first attempt then the search time $T$ and energy consumption $J$ are equivalent. On the other hand, if the search is interrupted at least once then $T$ will exceed $J$ by the amount of time spent in states $\mathbf{L}$ and $\mathbf{W}$. Therefore, the joint density of $T$ and $J$ can be obtained by accounting for the possibilities of locating the object in $1,2,~\cdots$ attempts while including the time spent in states $\mathbf{L}$ and $\mathbf{W}$ in $T$ but not in $J$.  Let $\phi(x,t)$ be the joint probability density of search time and energy consumption:  $\phi(x,t)dxdt=\Pr[x\leq J\leq x+dx,~t\leq T\leq t+dt]$. Since each attempt is independent of its predecessors, we have for $t\geq x$:
\begin{equation*}
\phi(x,t)= \gamma_d(t)\delta(t-x)+\int_0^x \gamma_\iota(y) \psi(t-x) \gamma_d(x-y) dy  + \cdots %
\end{equation*}
and $\phi(x,t) = 0$ for $t<x$. Evaluating the 2D LT of $\phi(x,t)$ yields:
\begin{equation}\label{phi}
\hat{\phi}(\xi,s)=\int_0^\infty\int_x^\infty\phi(x,t)e^{-st-\xi x}dt dx = \frac{\bar{\gamma}_d(s+\xi)}{1-\bar{\psi}(s) \bar{\gamma}_\iota(s+\xi)}.
\end{equation}
The LT of the pdf of the total search time is $\bar{g}(s) = \hat{\phi}(0,s)$. Inversion of the LT for small and large values of the real variable is performed by taking the limit of the corresponding Laplace variable as it tends to $\infty$ and $0$ yielding:
\begin{equation}\label{g-asympt}
g(t) \sim
\left\{
  \begin{array}{ll}
    \gamma_d(t)+\frac{\mu r}{\mu+r}\left[\Gamma_d(t) - \int_0^t \gamma_d(t-\tau) e^{-(\mu+r)\tau}d\tau \right],& t ~\textnormal{small}, \\
    \beta_T e^{-\beta_T t}, & t ~\textnormal{large},
  \end{array}
\right.
\end{equation}
where $\Gamma_d(t)=\int_0^t \gamma_d(\tau)d\tau$ and $\beta_T^{-1} = [\frac{1}{r}+\frac{1}{\mu}]~e^{(b+\sqrt{b^2+2c(\lambda+r)}) D/c}$.
Similarly the pdf of the energy consumption is $\bar{h}(\xi) = \hat{\phi}(\xi,0)$ which can be asymptotically inverted to:
\begin{equation}\label{h-asympt}
h(x) \sim \left\{
  \begin{array}{ll}
    \gamma_d(x) + (\lambda+r)\Gamma_d(x),  & x ~\textnormal{small}, \\
    \beta_J e^{-\beta_J x},     & x ~\textnormal{large},
  \end{array}
\right.
\end{equation}
where $\beta_J^{-1} = \frac{ e^{(b+\sqrt{b^2+2c(\lambda+r)})D/c}}{\lambda+r}$. Since it may not be possible to analytically invert $\bar{g}(s)$ and $\bar{h}(\xi)$ for all values of $t$ and $x$, we perform numerical inversion \cite{Hoog82} using MATLAB \cite{Hollenbeck98}. Fig.~\ref{gh} shows that the pdf of the search time and energy consumption using asymptotic and numerical inversion \cite{Hoog82} agree well for a wide range of delay and energy values.

\subsection{Asymptotic Formula for $k$ out of $N$ to be Successful in Large Time $B$} \label{asymptote1}

\begin{figure}[t]\centering
   \includegraphics[width=0.7\textwidth]{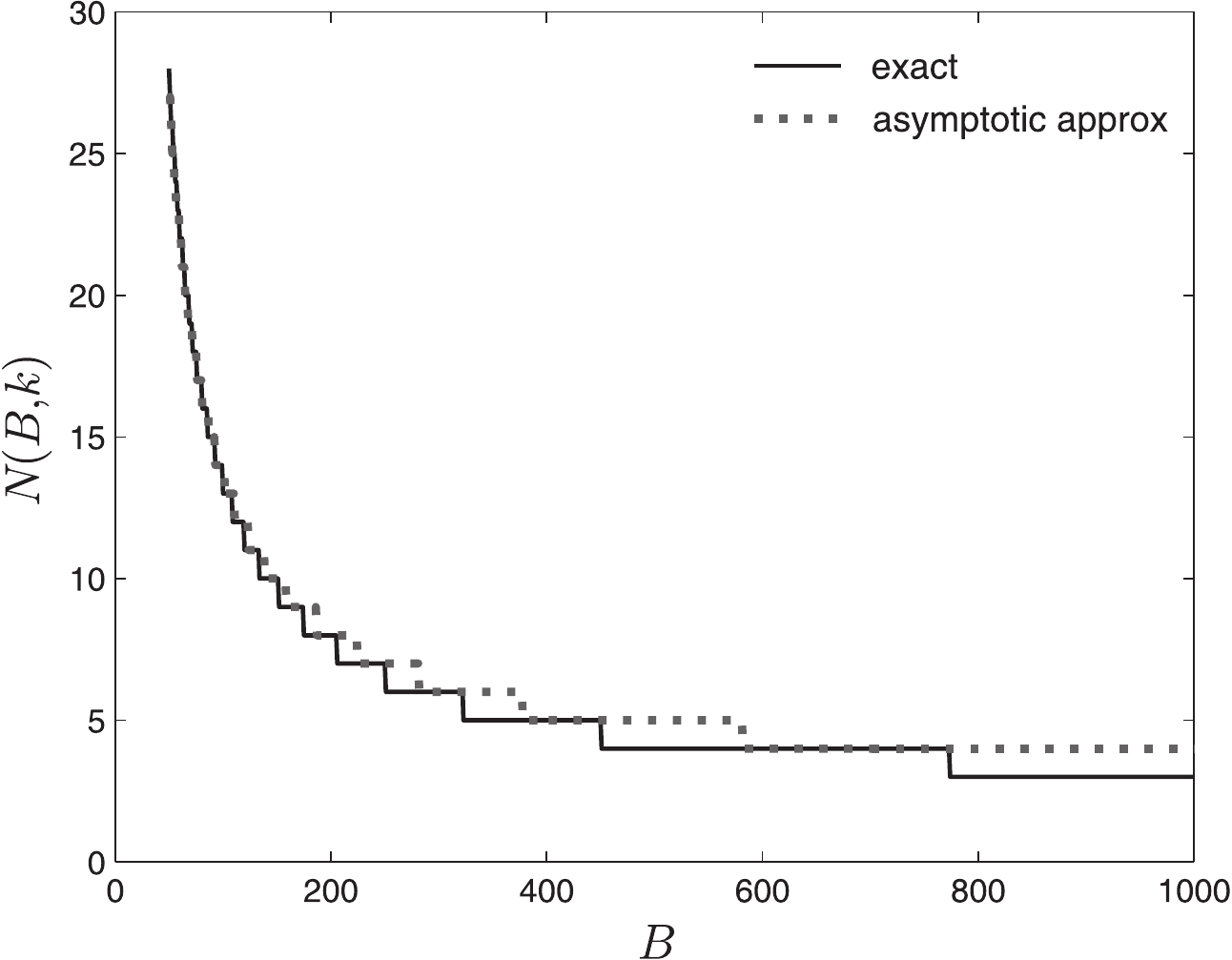}
   \caption{Comparison of the asymptotic approximation with exact analysis for the total number of searchers $N(B,k)$ that are required so that $k=3$ of them find the object within time $B$. Here $b=0$, $c=1$, $\lambda=0.0025$, $r^{-1} = 78$, $\mu^{-1}=10$ and $D=10$. } \label{noSearchers}
\end{figure}

The probability that $k$ out of $N$ independent searchers will be successful by time $t$ is:
\begin{equation}
G_{k,N}(t)\equiv \Pr[T_{k,N}\leq t] =\binom{N}{k}G(t)^k[1-G(t)]^{N-k}.\label{binom}
\end{equation}
Define $G^{-1}(p) = \inf\{t:~G(t) \geq p\},~ 0<p<1$, the {\em quantile function} of the distribution of the search time for a single searcher. When $N$ is large, it is known that $T_{\lceil pN \rceil,N}$, the $p$-th sample quantile, is asymptotically normally distributed \cite{sterfling80}:
\begin{equation}
T_{\lceil pN \rceil,N} \sim \mathcal{N}\left(G^{-1}(p),\frac{p(1-p)}{N~[g(G^{-1}(p))]^2} \right). \label{converge}
\end{equation}
Thus for large $N$ the distribution of the time for $k$ out of $N$ searchers to be successful tends to a constant equal to the $p\approx k/N$-th quantile of $G(t)$.
As a consequence, the number of searchers $N(B,k)$ required to find the object in time $B$ when $N$ is large is given approximately by:
\begin{equation}
N(B,k) \cong \left\lceil \frac{k}{G(B)} \right\rceil .\label{asymp}
\end{equation}
Since convergence to the normal distribution \eqref{converge} is fast, the expression \eqref{asymp} provides a good approximation even for relatively small $N(B,k)$. The good agreement between the asymptotic approximation of \eqref{asymp} and the
detailed analysis for $G_{k,N}(B)$ using the formula \eqref{binom} and the numerical inversion technique \cite{Hoog82} is illustrated in Fig.~\ref{noSearchers}.

\subsection{Energy Consumption} \label{energy}

In order to derive the energy required for $k$ out of $N$ independent searchers to locate the object, we need to evaluate the time-dependent pdf for the energy expended by a searcher. If we define $J(t)$ as the energy consumed by a searcher up to time $t$, then the joint pdf of $J(t)$ and the state of the searcher $s(t)$ can be evaluated in three distinct cases:

\noindent $\bullet$ The searcher reached the {\em destination} at some time $\tau\leq t$:
\begin{equation}
h_d(x,t) \equiv \frac{d}{dx}\Pr[J(t)\leq x, s(t) = \mathbf{P}] = \int_x^t \phi(x,\tau)d\tau, \quad \hat{h}_d(\xi,s) = \frac{\hat{\phi}(\xi,s)}{s}.
\end{equation}

\noindent $\bullet$ The searcher is {\em idle} at time $t$ due to a blocking or time-out that has not yet ended:
\begin{align*}
h_\iota(x,t)  \equiv&~ \frac{d}{dx}\Pr[J(t)\leq x, s(t) = \mathbf{W}~or~\mathbf{L} ]= \gamma_\iota(x)[1-\Psi(t-x)]\\
              +& \int_0^x\int_0^{t-x} \gamma_\iota(y_1) \psi(y_2) \gamma_\iota(x-y_1)[1-\Psi(t-x-y_2)]dy_2dy_1 +\cdots,
\end{align*}
where $\Psi(t)=\int_0^t\psi(\tau)d\tau$. Here the $i$-th term denotes the case where the time instant $t$ occurs after the search process is suspended $i$ times but before restarting the $(i+1)$-th search attempt. Taking the double LT of the above equation yields:
\begin{equation}
\hat{h}_\iota(\xi,s)= \frac{[1-\bar{\psi}(s)]\bar{\gamma}_\iota(s+\xi)}{s[1-\bar{\psi}(s)\bar{\gamma}_\iota(s+\xi)]}
                    = \frac{1}{s} \big[1- \frac{1-\bar{\gamma}_\iota(s+\xi)}{1-\bar{\psi}(s)\bar{\gamma}_\iota(s+\xi)} \big].
\end{equation}

\noindent $\bullet$ The searcher is {\em active} at time $t$:
\begin{align*}
h_a(x,t) \equiv ~& \frac{d}{dx}\Pr[J(t)\leq x, s(t) = \mathbf{S}] = e^{-(\lambda+r)t}[1-G_0(t)]\delta(x-t) \\
           &  +\int_0^x \gamma_\iota(y) \psi(t-x) e^{-(\lambda+r)(x-y)}[1-G_0(x-y)] dy+\cdots.
\end{align*}
In the first term, no time-out or blocking has occurred up to $t$ and consequently the total energy consumption is equal to $t$. The $i$-th term
corresponds to the case where at time $t$ the search is ongoing after it was restarted $i-1$ times so that the pdf of the energy utilisation up to $t$ is given by the convolution of the pdf of $i-1$ interrupted search periods (each followed by an idle period in which energy is not consumed) and a single search period which does not end before the time instant $t$. We then end up with:
\begin{equation}
\hat{h}_a(\xi,s) =  \frac{1-\bar{g}_0(s+\xi+\lambda+r)}{[s+\xi+\lambda+r][1-\bar{\psi}(s)\bar{\gamma}_\iota(s+\xi)]}=  \frac{1}{\lambda+r} \frac{\bar{\gamma}_\iota(s+\xi)}{1-\bar{\psi}(s)\bar{\gamma}_\iota(s+\xi)}.
\end{equation}
Note that $\sum_{p=\{d,\iota,a\}}h_p(x,t)\equiv h(x,t)$ is the pdf of the energy consumed up to $t$ by the searcher $\hat{h}(\xi,s) = \frac{1}{s} [1 - \frac{\xi}{\lambda+r} \frac{\bar{\gamma}_\iota(s+\xi)}{1-\bar{\psi}(s)\bar{\gamma}_\iota(s+\xi)}]$ and $\lim_{t\to \infty} h(x,t) = h(x)$ as expected.

On the other hand, if all active but unsuccessful searchers continue searching after the first $k$ successful ones complete their search, we need to know the energy expended by an active searcher up to $t$, and its distance to the object being sought so as to compute the additional energy consumed before it actually stops moving. Define $f(z,x,t)dz dx  = \Pr[x\leq J(t)\leq x+dx, z\leq Z(t)\leq z+dz]$ where $x\leq t, z> 0$, which can be derived using $f_0(z,t)$ from \eqref{f0-sol} as:
\begin{align*}
f(z,x,t) =~& e^{-(\lambda+r)t} f_0(z,t)\delta(x-t) \\
           & + \int_0^x \gamma_\iota(\tau)\psi(t-x) f_0(z,x-\tau)e^{-(\lambda+r)(x-\tau)} d\tau +\ldots,
\end{align*}
where the first term is the probability that the searcher reaches distance $z$ in time $t$ without being interrupted so that $x=t$; the second term is the probability that the search is stopped at some time $\tau\in[0,x]$, it is restarted after $t-x$ time units, and distance $z$ is reached during the next search attempt in a time interval $x-\tau$ in which no interruption occurs. More generally, the $i$-th term represents the case where the time instant $t$ lies in the $i$-th attempt to locate the object, when the searcher is at distance $z$ and has consumed $x$ units of energy. The 2D LT of $f(z,x,t)$ is:
\begin{equation}
\hat{f}(z,\xi,s) = \frac{\bar{f}_0(z,s+\xi+\lambda+r)}{1-\bar{\psi}(s)\bar{\gamma}_\iota(s+\xi)}.
\end{equation}
Note that $f(z,x,t)$ satisfies the equalities $\int_0^t f(z,x,t) dx = f(z,t)$ and $\int_0^\infty f(z,x,t) dz = h_a(x,t)$ so that $\bar{f}(z,s) = \hat{f}(z,0,s) = \frac{\bar{f}_0(z,s+\lambda+r)}{1-\bar{\psi}(s)\bar{\gamma}_\iota(s)}$.

Let $J_{k,N}^{~-}$ and $J_{k,N}^+$ be the total energy consumption when search is suspended immediately after the object being sought is found by $k$ searchers, and when there is no mechanism to immediately stop active searchers after the completion of the search, respectively. We then have:

\begin{result}\label{result:hkN}
The LT of the pdf for the total energy consumption is:
\begin{align}\nonumber
&\bar{h}_{k,N}^\pm(\xi) \equiv \frac{d}{dx}\Pr[J_{k,N}^\pm\leq x] \\\label{hkN}
&\quad                            = \frac{N!}{(k-1)!(N-k)!} \int_0^\infty \bar{h}_d(\xi,t)^{k-1} \bar{\phi}(\xi,t) [\bar{h}_\iota(\xi,t) + \bar{h}_\ast(\xi,t)]^{N-k} dt,
\end{align}
where $h_\ast$ is substituted by $h_a$ in the case of $h_{k,N}^{-}$ while for $h_{k,N}^{+}$  it is replaced by the pdf of the total energy expended by an unsuccessful active searcher until it stops moving after the search ends at $t$:
\begin{align*}
h_c(x,t) =& \int_0^\infty \int_0^x f(z,x-u,t) \big[\gamma_\iota(u|z) + \gamma_d(u|z)\big] du dz,\\
\hat{h}_c(\xi,s) =& \frac{1}{s(\xi+\lambda+r)} ~ \frac{s\bar{\gamma}_\iota(s+\xi) + \xi[\bar{\gamma}_d(\xi)-\bar{\gamma}_d(s+\xi)]}{1-\bar{\psi}(s)\bar{\gamma}_\iota(s+\xi)},
\end{align*}
and $\gamma_\iota(.|z)$ and $\gamma_d(.|z)$ are computed as in \eqref{gamma-interr} and  \eqref{gamma-dest}, respectively, but with initial distance $z$ instead of $D$.
\end{result}
\begin{proof}
For $J_{k,N}^-$ to be equal to $x$ with a search time $t$, it is necessary that exactly $k - 1$, $1$ and $N - k$ searchers locate the object being sought in the intervals $[0,t]$, $[t,t+dt]$ and $[t+dt,\infty]$ respectively, and that the energy expended by each individual searcher is at most $t$ while their sum is $x$. The probabilities that a search succeeds in the three respective intervals while consuming $w$ units of energy up to $t$ are $h_d(w,t)dw$, $\phi(w,t)dwdt$ and $[h_\iota(w,t)+h_a(w,t)]dw$.
$h_{k,N}^-(x)$ then follows by accounting for all possible combinations, convolving with respect to the energy variable (which is equivalent to multiplication in the $\xi$-domain), and integrating over all possible values of $t$. The pdf $h_{k,N}^+$ is evaluated in the same manner as $h_{k,N}^-$ except that  we use $h_c$ instead of $h_a$ to take account of the additional energy consumed by active searchers upon the completion of the search. Specifically, if the object being sought is found at some time $t$ while a searcher is at distance $z>0$ and has consumed $x-u$ units of energy, then the searcher will continue to move and consume additional $u$ units of energy with probability $\gamma_\iota(u|z)du$ if it is interrupted before reaching the destination or $\gamma_d(u|z)du$ otherwise. $h_c(y,t)$ is obtained by integrating over all possible values of the distance $z> 0$ and energy $u\in[0,x]$ at the time instant $t$, and its double LT follows directly.
\end{proof}

\begin{figure}[t]\centering
   \includegraphics[width=0.7\textwidth]{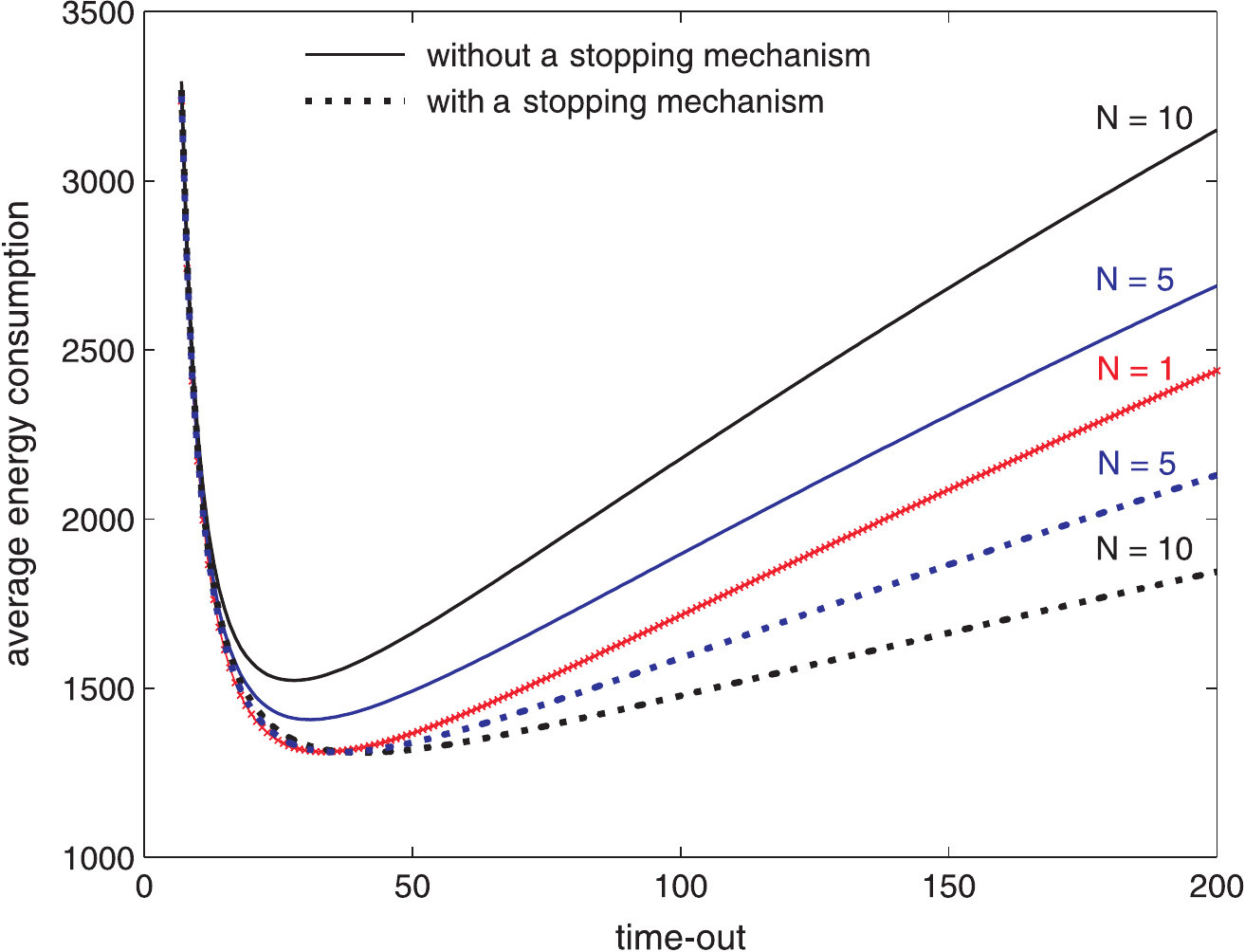}
   \caption{Average energy consumption (with and without a stopping mechanism) versus time-out $1/r$ for $k=1, b=0.15, c=1.25, \lambda=0.001, \mu=0.1, D=10$ and different values of $N$. When $N$ increases, the {\em energy consumed increases if there is no stopping mechanism}, and the opposite is true with the stopping mechanism.}\label{timeout}
\end{figure}

The total average energy consumption is then $E[J_{k,N}^\pm] = -\lim_{\xi\to 0}\frac{d \bar{h}_{k,N}^\pm(\xi)}{d \xi}$ yielding:
\begin{align}\nonumber
E[J_{k,N}^\pm] =& - \lim_{\xi\to 0}\frac{N!}{(k-1)! (N-k)!}\int_0^\infty dt\bigg[G(t)^{k-1}[1-G(t)]^{N-k} \frac{\partial \bar{\phi}(\xi,t)}{\partial \xi} \\\nonumber
& + (k-1)G(t)^{k-2}g(t)[1-G(t)]^{N-k}\frac{\partial \bar{h}_d(\xi,t)}{\partial \xi} +(N-k)  \\
&\times G(t)^{k-1}g(t)[1-G(t)]^{N-k-1}[\frac{\partial\bar{h}_\iota(\xi,t)}{\partial\xi}+\frac{\partial\bar{h}_\ast(\xi,t)}{\partial\xi}]\bigg],
\end{align}
which, unlike $h_{k,N}^\pm(x)$, requires only 1D LT numerical inversions for $\frac{\partial \bar{\phi}(\xi,t)}{\partial \xi}$, $\frac{\partial \bar{h}_d(\xi,t)}{\partial \xi}$ and $\frac{\partial}{\partial \xi}[\bar{h}_\iota(\xi,t)+\bar{h}_\ast(\xi,t)]$ when $\xi \to 0$. Fig.~\ref{timeout} shows  $E[J_{k,N}^-]$ and $E[J_{k,N}^+]$  versus the time-out $1/r$ for different values of $N$. One sees that the minimum energy consumed until the object is found (i.e. $E[J_{k,N}^-]$) does not vary much with the number of searchers $N$. However, in the absence of a stopping mechanism the minimum energy consumed by the search {\em increases} with $N$ while the ``optimum'' time-out decreases in order to reduce the additional energy wasted by active searchers after the completion of the search. Of course the intuitive but interesting observation is that when $N$ increases, the {\em energy consumed increases if there is no
stopping mechanism}, while the opposite occurs with the stopping mechanism.

\section{Search in a Non-homogeneous Medium}\label{non-hom}

This section considers a {\em single} searcher in an infinite random non-homogeneous medium, with spatially non-homogeneous events that may stop or impede the current search \cite{per2011,CJ-diffusion2012}. A motivation for this work is the case where the object being sought is protected from the searcher: as the searcher approaches, its progress becomes more frequently blocked or destroyed, and a new searcher has to be sent out to replace it. Another example of a non-homogeneous search space occurs when the search progresses faster as the searcher approaches the object, for instance when directional information (the smell of food for the forager or a radio signature in a wireless network) becomes more readily available in the proximity of the object. We develop an analytical solution technique based on a finite but unbounded number of internally homogeneous segments, yielding the average search time and the energy expended. The results are illustrated by several examples.

We simplify the model of a non-homogeneous search space, described in \eqref{one}$-$\eqref{four} with $N=1$ and $a=0$, by considering a finite but unbounded number of ``segments'' that have different parameters for the Brownian motion describing the searcher's movement as a function of its distance to the object being sought, while within each segment the parameters are the same. The first segment is in the immediate proximity of the object being sought, starting at distance $z=0$. Each segment may have a different size, and we assume that there are a total of $m<\infty$ segments. By choosing as many segments as we wish, and letting each segment be as small as we wish (all segments need not be of the same length), we can approximate as closely as needed any physical situation that arises where the searcher's motion characteristics vary over the distance of the searcher to the object being sought. We also show that this discrete representation leads to a neat algebraic ``product form'' representation of the average search time, and that it thus provides a useful analytic form that offers a more intuitive representation of the analytical results.

\begin{figure}[t]\centering
  \includegraphics[width=0.95 \textwidth]{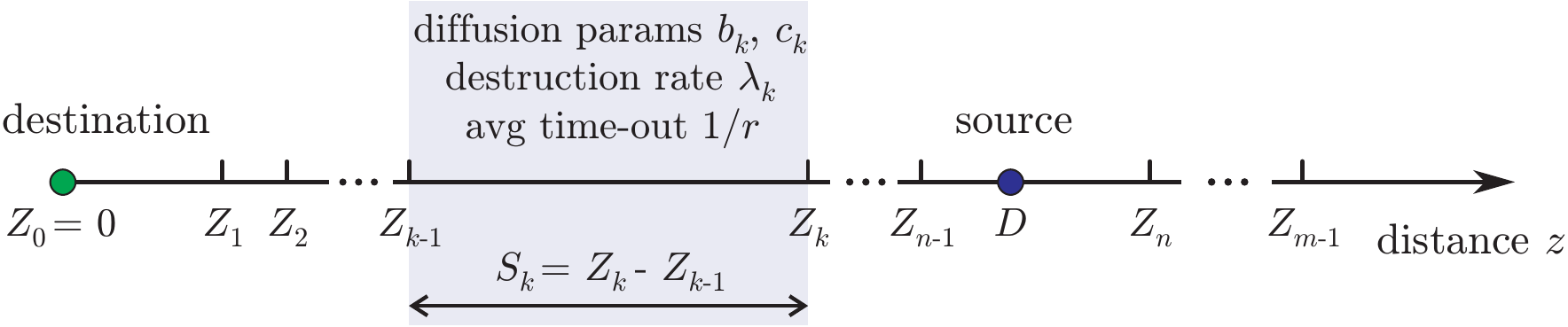}
  \caption{Piece-wise approximation of a non-homogeneous medium.}\label{illustration}
\end{figure}

We denote by $0\leq Z_k<\infty$ the boundary between the $k$-th and $(k+1)$-th segments with $Z_0 = 0$. The last segment goes from $Z_{m-1}$ to $+\infty$, and we assume that both $m$ and $Z_{m-1}$ are finite but unbounded. Thus for greater accuracy in representing the search we can take as many segments as we wish, and they may be as small as needed, but they are all finite except the last segment. Thus for $0\leq k \leq m$, the $k$-th segment represents the range of distances $Z_{k-1}\leq z <Z_k$, and let $S_k = Z_k - Z_{k-1}$ denote its size. We use $n$ to denote the segment number in which the source point of the search is located, i.e. $Z_{n-1}\leq D < Z_n$. The piece-wise approximation is illustrated in Figure~\ref{illustration}. If we write the parameters of the diffusion model for the $k$-th segment as:
\begin{equation*}
\{f(z,t),~b(z),~c(z),~\lambda(z)\} = \{f_k(z,t),~b_k,~c_k,~\lambda_k\}, \quad  Z_{k-1}\leq z < Z_k,
\end{equation*}
then the differential equation for the {\em stationary} solution of the distance dependent diffusion equation for $k \neq n$ is:
\begin{equation*}
0 =\frac{c_k}{2}\frac{d^2f_k(z)}{dz^2}-b_k\frac{df_k(z)}{dz}-(\lambda_k+r)f_k(z),
\end{equation*}
while the equation for the segment where the source is located is:
\begin{equation*}
-[P+\mu  W]\delta(z-D) =  \frac{c_n}{2}\frac{d^2f_n(z)}{dz^2}-b_n\frac{df_n(z)}{dz}-(\lambda_n+r)f_n(z).
\end{equation*}
We will also have:
\begin{align*}
rL    &= \sum_{k=1}^m \lambda_k \int_{Z_{k-1}}^{Z_k} f_k(z)dz,\\%
\mu W &= r [L +  \sum_{k=1}^m \int_{Z_{k-1}}^{Z_k} f_k(z)dz],\\%
P     &= \lim_{z \rightarrow 0^+}[\frac{c_1}{2}\frac{df_1(z)}{dz}-b_1 f_1(z)],%
\end{align*}
and the normalisation condition:
\begin{equation*}
1=P+W+L+\sum_{k=1}^m \int_{Z_{k-1}}^{Z_k} f_k(z) dz.
\end{equation*}

\begin{result}
The total average search time, which is obtained by solving for $P$ so that $E[T]=P^{-1}-1$, is given by:
\begin{equation}\label{Avg}
E[T]=\big(\frac{1}{r}+\frac{1}{\mu}\big) \big[\sqrt{\frac{b_n^2+2c_n(\lambda_n+r)}{b_1^2+2c_1(\lambda_1+r)}}
\frac{{A}_n {E}_n e^{u_n S_n}-{B}_n {C}_n e^{v_n S_n}}{{E}_n e^{u_n (Z_n - D)}+{C}_n e^{v_n (Z_n - D)}}-1\big],
\end{equation}
where $u_k,v_k=\frac{b_k\pm \sqrt{b_k^2+2c_k(\lambda_k+r)}}{c_k}$ and the remaining parameters in \eqref{Avg} are computed as follows. Define:
\begin{align*}
\alpha_k^- &=\frac{c_k u_k - c_{k-1} v_{k-1}}{c_k(u_k - v_k)},\quad \beta_k^-  =\frac{c_k u_k - c_{k-1} u_{k-1}}{c_k(u_k - v_k)},\\%
\alpha_k^+ &= \frac{c_k u_k-c_{k+1} v_{k+1}}{c_k(u_k-v_k)},\quad \beta_k^+  = \frac{c_k u_k-c_{k+1} u_{k+1}}{c_k(u_k-v_k)}.%
\end{align*}
Then set ${A}_1=1$ and ${B}_1=-1$ and for $2\leq k \leq n$ compute:
\begin{equation*}
\left[\begin{array}{c}
  {A}_{k}  \\
  {B}_{k}
\end{array}\right]=
\left[\begin{array}{cc}
  \alpha_{k}^- & \beta_{k}^- \\
  1-\alpha_{k}^- & 1-\beta_{k}^-
\end{array}\right]%
\left[{\setlength\arraycolsep{0.1em}\begin{array}{cc}
  e^{u_{k-1} S_{k-1}} & 0 \\
  0 & e^{v_{k-1} S_{k-1}}
\end{array}}\right]
\left[\begin{array}{c}
  {A}_{k-1}  \\
  {B}_{k-1}
\end{array}\right].
\end{equation*}
Then set  ${C}_m=0$ and ${E}_m=e^{v_m Z_m}$, and start another computation at $k=m-1$  for  $n\leq k \leq m-1$ with:
\begin{equation*}
\left[\begin{array}{c}
  {C}_{k}  \\
  {E}_{k}
\end{array}\right]=%
\left[\begin{array}{cc}
  \alpha_{k}^+ & \beta_{k}^+ \\
  1-\alpha_{k}^+ & 1-\beta_{k}^+
\end{array}\right]%
\left[{\setlength\arraycolsep{0.1em}\begin{array}{cc}
  e^{-u_{k+1} S_{k+1}} & 0 \\
  0 & e^{-v_{k+1} S_{k+1}}
\end{array}}\right]
\left[{\setlength\arraycolsep{0.1em}\begin{array}{c}
  {C}_{k+1}  \\
  {E}_{k+1}
\end{array}}\right].
\end{equation*}
\end{result}
This completes all terms in $E[T]$, and the proof \cite{CJ-diffusion2012} consists in showing that the pdf $f_k(z)$ has the form:
\begin{equation*}
f_k(z)=\left\{
\begin{array}{ll}
\eta [{A}_k e^{u_k(z-Z_{k-1})} + {B}_k e^{v_k(z-Z_{k-1})}],& Z_{k-1}\leq z\leq \min(D,Z_k), \\
\sigma [{C}_k e^{-u_k(Z_k-z)} + {E}_k e^{-v_k(Z_k-z)}],& \max(D,Z_{k-1})\leq z\leq Z_k,
\end{array}\right.
\end{equation*}
where $\eta   = \omega [{E}_n e^{u_n(Z_n-D)} + {C}_n e^{v_n(Z_n-D)}]$, $\sigma = \omega  [{A}_n e^{u_n S_n}e^{v_n (Z_n-D)} + {B}_n e^{v_n S_n}e^{u_n (Z_n-D)} ]$ and
\begin{align*}
\omega =&\frac{r\mu/(r+\mu)}{\sqrt{b_n^2+2c_n(\lambda_n+r)}} \big\{ {A}_n E_n e^{u_n S_n} - {B}_n C_n e^{v_n S_n} \\
        & -[1-\frac{r\mu}{r+\mu}]\sqrt{\frac{b_1^2+2c_1(\lambda_1+r)}{b_n^2+2c_n(\lambda_n+r)}} [E_n e^{u_n(Z_n-D)} + C_n e^{v_n(Z_n-D)}]\big\}^{-1}.
\end{align*}
Note that if the last segment includes the starting point $z=D$ (i.e. $m=n$) then $E[T]$ takes the much simpler form:
\begin{equation}\label{m=n}
E[T]=\frac{r+\mu}{r\mu} [\sqrt{\frac{b_n^2+2c_n(\lambda_n+r)}{b_1^2+2c_1(\lambda_1+r)}} {A}_n e^{u_n (D-Z_{n-1})}-1],
\end{equation}
and with a homogeneous search space $m=n=1$ so that $E[T]=\frac{r+\mu}{r\mu}[e^{u_1 D}-1]$ as we would expect from \eqref{ET1N} when $N=1$.

\subsection{Search in a Protected Neighbourhood} \label{protect}

\begin{figure}[t]
        \centering
        \begin{subfigure}[t]{0.49\textwidth}\centering
                \includegraphics[width=\textwidth]{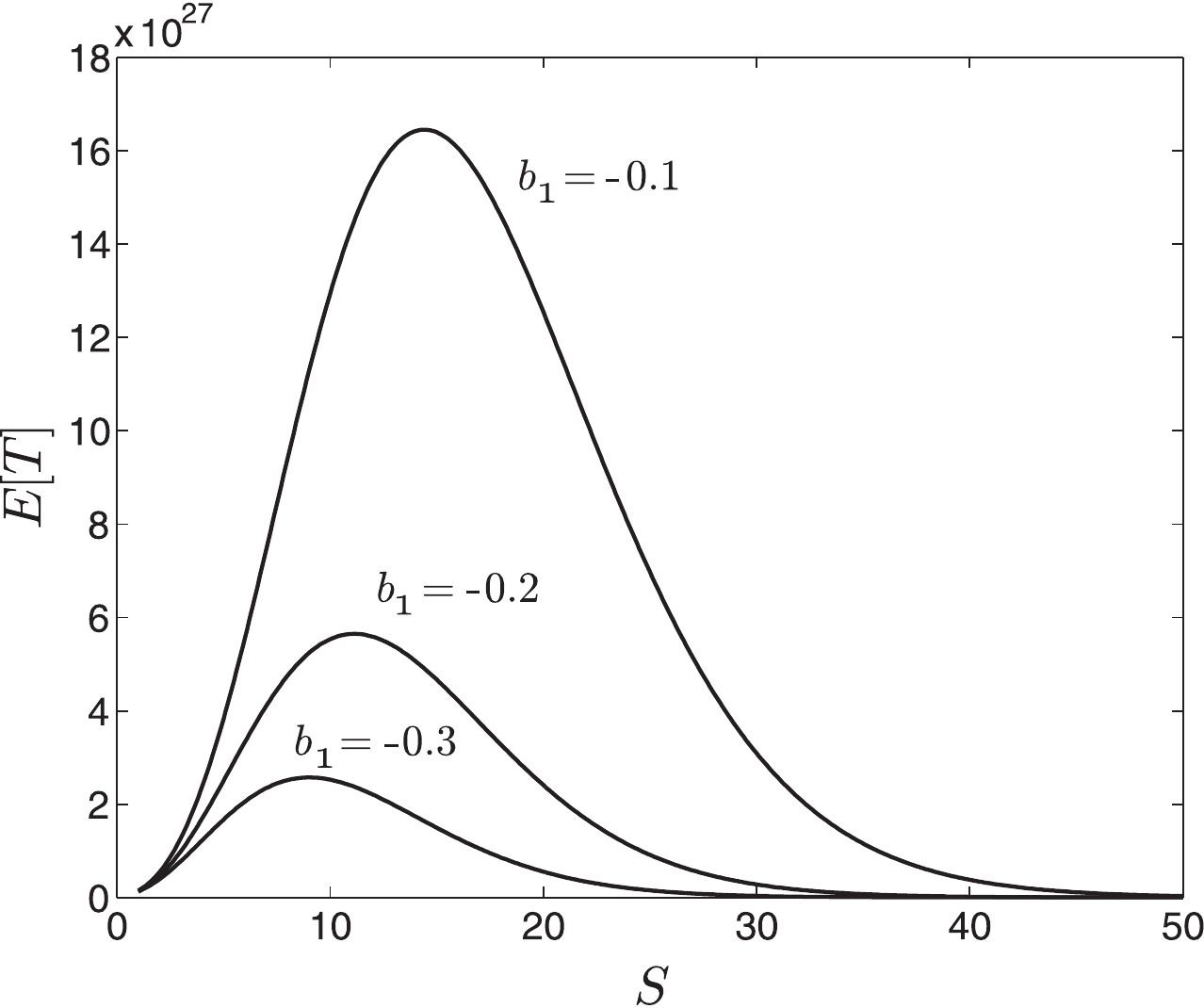}\caption{}\label{prot-a}
        \end{subfigure}~
        \begin{subfigure}[t]{0.49\textwidth}\centering
                \includegraphics[width=\textwidth]{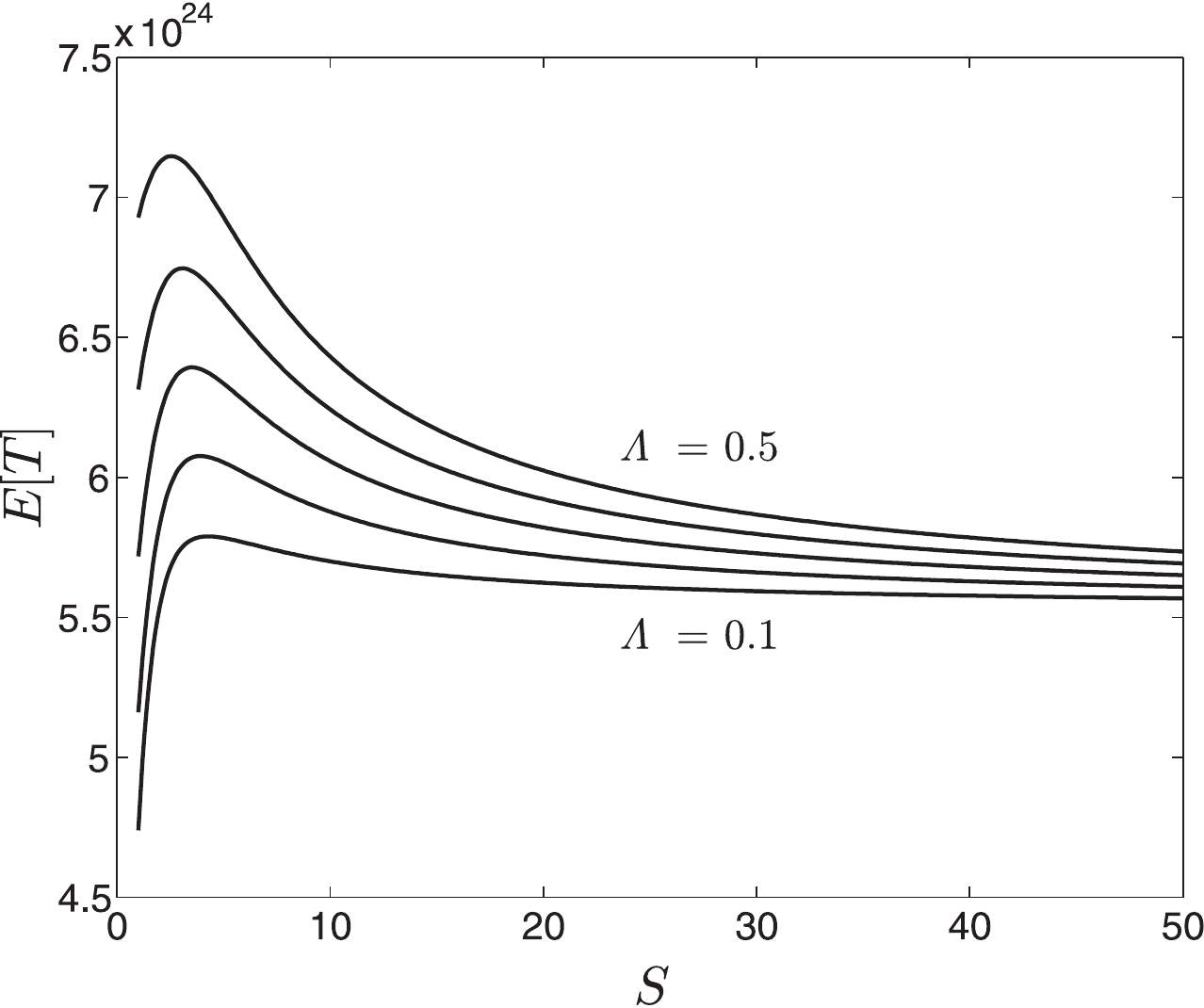}\caption{}\label{prot-b}
        \end{subfigure}
        \caption{Average search time $E[T]$ versus size of the protected neighbourhood $S$ when (a) $\lambda_1 = 10/S$ for different values of $b_1$: the optimum protection area needed becomes smaller so that $\lambda_1$ increases when the search speed increases (b) $\lambda_1 = \Lambda/S^2$ for $\Lambda = 0.1$ to $0.5$ with a step of size 0.1 and $b_1 = b_2 = 0.25$: the protection area needed to maximise the search time decreases as $\Lambda$ increases.}\label{prot}
\end{figure}

We consider the case where the neighbourhood of the object  being sought, up to a distance $S$, is protected by randomly located traps that destroy the searcher. In the rest of the search space accidental destruction of the searcher may occur, but at much lower rate. Thus we take $m=n=2$, so that $E[T]$ is obtained from
\eqref{m=n} with $\lambda_1>>\lambda_2$. We choose the time-out parameter $r$ so as to minimise $E[T]$ when $D=100$, $b_k=0.25$, $c_k = 1$, $\lambda_k=0$, $\mu=0.1$ and $S=10$. In Fig. \ref{prot} we raise the question about how to select $S$ and $\lambda_1$ together in order to maximise the protection offered to the object being sought. Thus we take $\lambda_1$ to be inversely proportional to $S$ in Fig.~\ref{prot-a} so that the average number of sources of protection, placed at rate $\lambda_1$, remains constant in proportion to the protection space of size $S$. The mapping of time rate to spatial rate will remain constant for any fixed value of $b_1$ which is the speed of motion. In this context, we examine whether there is a size $S^*$ of the protected neighbourhood which maximises protection, i.e. that maximises the average time to locate the object. Fig.~\ref{prot-a} shows that there is indeed an optimum $S^*$ that varies with the speed $b_1$ of the searcher inside the protected neighbourhood. As the speed increases, the optimum size of the neighbourhood gets smaller: a smaller size implies a higher ``rate of protection'' and hence more frequently occurring destructions of the searcher which compensate for the higher speed of the searcher. However the corresponding maximum values of $E[T]$ do become smaller as the searcher's speed increases. In Fig.~\ref{prot-b} we set $b_1=b_2=0.25$ and $\Lambda$ is varied in $\lambda_1 = \Lambda/S^2$. The results are similar to the previous ones.

\subsection{A Phase Transition Effect} \label{phase}

\begin{figure}[t]\centering
\includegraphics[width=0.7 \textwidth] {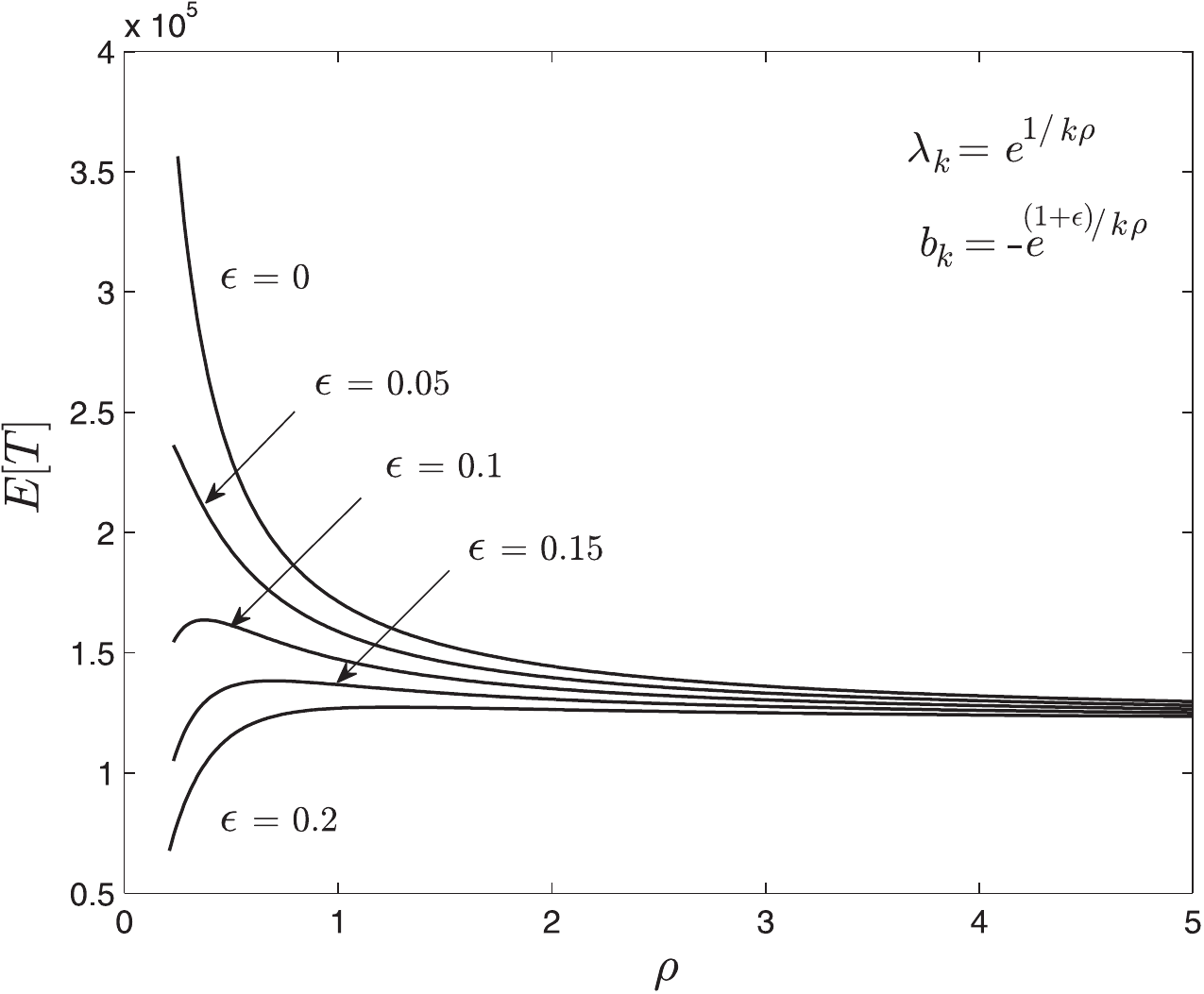}
\caption{ \label{fig7} Average search time $E[T]$ versus $\rho$ when $\lambda_k = e^{\frac{1}{k\rho}}$ and $b_k = - e^{\frac{1+\epsilon}{k\rho}}$ for different values of $\epsilon\geq 0$; $c_k=1$, $D=10$, $r=0.05$, $\mu=0.025$ and $S_k = 1$ for $k < m=20$.}
\end{figure}

The destruction of the searcher and the time-out will both relaunch the search process allowing the searcher to improve its chances to attain the object, we suspect that if the object being sought is heavily defended when the searcher gets very close to it, then the searcher may never attain the object. This is confirmed in Fig.~\ref{fig7} where we observe that if $\lambda_k=e^{\frac{1}{k\rho}}$ and $b_k = -e^{\frac{1+\epsilon}{k\rho}}$ with $\epsilon \geq 0$, then as $\rho$ becomes very small, $E[T]$ tends to infinity despite the fact that near the origin the search speed is greater and its randomness is smaller. However it is interesting to see that if the searcher's speed of approach to the object grows faster than the rate at which the searcher may be destroyed, then both $E[T]$ remains finite and may tend to zero, while in the opposite case it will tend to infinity presenting a form of phase transition.

\section{Future Work}

In future work, we expect to address issues of load sharing or balancing so as to achieve overall better performance in search activities \cite{Aguilar}. For instance, it would be reasonable to subdivide the search space among multiple searchers so that each search space is covered by a subset of the searchers. Also, it may be interesting to investigate how it would be possible to exploit the distinct performance capacities of multiple classes of searchers \cite{Multiple} so as to improve overall performance with respect to both time and energy needed for a search.

Another interesting area of research concerns the analysis of searchers that learn from each other so that they take advantage of their more successful colleagues. Similarly we can evaluate situations where searchers may try to conceal their knowledge of the location of an object, or even deceive the other searchers so as to impede their success.
Thus we see that this area of research still reserves many opportunities for interesting research problems that are of value to different areas of application.


\end{document}